# Detecting Linear Dichroism with Atomic Resolution

Roger Guzman[1,+], Ján Rusz[2], Ang Li[1], Juan Carlos Idrobo[3,4,*], Wu Zhou[1,*], Jaume Gazquez[5,*]

[1]School of Physical Sciences, University of Chinese Academy of Sciences, Beijing, China.

[2]Department of Physics and Astronomy, Uppsala University, Uppsala, Sweden.

[3]Department of Materials Science and Engineering, University of Washington, Seattle, WA, USA.

[4]Physical and Computational Sciences Directorate, Pacific Northwest National Laboratory, Richland, WA

[5]Institute of Materials Science of Barcelona (ICMAB-CSIC), Bellaterra, Barcelona, Spain.

*Corresponding authors: (JG) jgazquez@icmab.es; (WZ) wuzhou@ucas.ac.cn; (JCI) jidrobo@uw.edu

[+]Present address:

Institute of Materials Science of Barcelona (ICMAB-CSIC), Bellaterra, Barcelona, Spain.

**ABSTRACT**

**X-ray linear dichroism has been pivotal for probing electronic anisotropies, but its inherent limited spatial resolution precludes atomic-scale investigations of orbital polarization. Here we introduce a versatile electron linear dichroism methodology in scanning transmission electron microscopy that overcomes these constraints. By exploiting momentum-transfer-dependent electron energy-loss spectroscopy with an atomic-sized probe, we directly visualize orbital occupation at individual atomic columns in real space. Using strained $La_{0.7}Sr_{0.3}MnO_3$ thin films as a model system, we resolve the Mn-3$d$ $e_g$ orbital polarization with sub-angstrom precision. We show that compressive strain stabilizes $3z^2$-$r^2$ occupation while tensile strain favors $x^2$-$y^2$. These results validate our approach against established X-ray measurements while achieving the ultimate single atomic-column sensitivity. We further demonstrate two optimized signal extraction protocols that adapt to experimental constraints without compromising sensitivity. This generalizable platform opens unprecedented opportunities to study symmetry-breaking phenomena at individual defects, interfaces, and in quantum materials where atomic-scale electronic anisotropy governs emergent functionality.**

Linear dichroism measurements provide insights into the anisotropic electronic structure of materials, directly probing orbital occupation and charge ordering [1–4]. This high sensitivity to direction-dependent electronic responses makes the technique versatile for investigating emergent quantum phenomena and interfacial effects in low-dimensional systems and thin-film heterostructures[3,4].

Analogous to X-ray linear dichroism (XLD), electron linear dichroism (ELD) exploits the momentum transfer vector $\hbar \mathbf{q}$ as the electron scattering counterpart to the X-ray polarization vector, with $\mathbf{q} = \mathbf{k_i} - \mathbf{k_f}$, for incident and scattered wavevectors $k_i$ and $k_f$, respectively. In ELD, the dichroic contrast arises from the angular dependence of inelastic scattering on the orientation between $\mathbf{q}$ and the crystal reference frame defined by position vector $\mathbf{r}$ (which encodes the directionality of the unoccupied electronic states), as described by the dipole approximation in the double differential scattering cross section (DDSCS) [5]. This principle has enabled multiple approaches to record $\mathbf{q}$-dependent electron energy loss spectroscopy (EELS) to study electronic anisotropies in (S)TEM[6–17], but without achieving atomic resolution.

Here, we present a straightforward $\mathbf{q}$-dependent EELS spectrum imaging methodology that directly extracts atomic-resolution ELD signals in real space using an atomic-sized STEM probe. Compatible with any aberration-corrected STEM, this approach provides the unique capability to map orbital polarization at individual atomic columns when sufficient signal-to-noise ratio (SNR) is achieved. This technique unlocks unprecedented capabilities for probing electronic anisotropies at previously inaccessible scales —from heterostructures and interfaces down to single-atom defects— offering unique insights beyond the reach of any existing methodology.

We demonstrate the ELD application in $La_{0.7}Sr_{0.3}MnO_3$ (LSMO) thin films, a prototypical mixed-valence manganite where strain-mediated orbital polarization is well-established[18–20]. In LSMO thin films grown on (001)-mismatched substrates, tetragonal distortion modifies the out-of-plane $c$ lattice parameter, distorting $MnO_6$ octahedra. This symmetry breaking lifts the degeneracy of the $Mn^{3+}$ $3d$ $e_g$ orbitals ($x^2$-$y^2$ and $3z^2$-$r^2$). Specifically, tensile strain favors $x^2$-$y^2$ occupancy, whereas compressive strain stabilizes $3z^2$-$r^2$ (see **Fig. 1a** and **Extended Data Fig. 1**), profoundly affecting the transport and magnetic properties of LSMO thin films[21].

In a STEM-EELS experiment using a forward-scattering geometry, the momentum transfer vector **q** can be decomposed into components parallel ($q_∥$) and perpendicular ($q_⊥$) to the incident wave vector $k_i$. When the incident beam is oriented along the [010] zone axis of the $MnO_6$ octahedra, the $x^2$-$y^2$ and $3z^2$-$r^2$ orbitals align with the in-plane ($x, y$) and out-of-plane ($z$) crystallographic axes, respectively (**Fig. 1a**). The matrix element for an electronic transition to these orbitals maximizes when the scalar product **q · r** is largest, that is, when $q_⊥$ aligns with the spatial extension of the orbitals [22]. Consequently, $q_{⊥xy}$ selectively probes the $x^2$-$y^2$ orbital states (in-plane anisotropy), while $q_{⊥z}$ the $3z^2$-$r^2$ states (out-of-plane anisotropy). This directional sensitivity enables direct mapping of orbital polarization by resolving momentum-dependent energy loss near edge structure (ELNES) variations in anisotropic materials.

By performing a STEM raster scan centered on Mn atomic columns (**Fig. 1b**), the ELD signal from the Mn-$L_{2,3}$ edge can be retrieved. Here, the ELD is defined as the difference in total intensity between horizontal and vertical scan directions, $SI_h$ and $SI_v$ respectively (**Fig. 1c**), which are dominated by $q_{⊥x}$ and $q_{⊥z}$ momentum transfers, i.e. $R_x$ and $R_z$ beam scan positions (blue and red boxes in **Fig. 1b**). These momentum components directly probe anisotropy of the unoccupied Mn $3d$ states. **Figure 1d** shows the EELS spectra from the pixels along the $x$-scan line and $z$-scan line, respectively. In the dipole regime[5], and under these experimental conditions, the spatial variation of **r** dominates over **q**, so the ELD contrast scales with the displacement ($±Δr$) from the Mn atomic center (see **Extended Data Fig. 2**). A straightforward extraction method integrates the total intensity from the $R_x$ and $R_z$ beam positions (pixels) along the two scan directions (line-integration method), yielding $I_h$ and $I_v$ respectively (**Fig. 1e**). The ELD signal is calculated as $I_h - I_v$, as schematically illustrated in **Fig. 1f**.

However, the lobed spatial distribution of $d$-orbitals extends beyond the linear scan paths described in **Fig. 1c**, causing the line-integration method to underestimate the ELD signal. This underestimation arises because the radial extent of **r**, weighted by the $d$-orbital wavefunctions, contributes to dichroism at beam positions displaced vertically (horizontally) from the $x$ ($z$) scan axis. While the linear scans provide an intuitive experimental framework, the difference $I_h - I_v$ includes partial signal cancellation from the central region of the atomic column. Although these central pixels contribute little to the dichroic contrast, they add noise to the differential signal. An improved strategy can be devised based on dynamical diffraction calculations showing intensity maps of the spatially resolved, direction-dependent scattering coefficients $n_{xx}$ and $n_{zz}$ (**Figs. 1g-h**), and the relative L-edge ELD signal for a Mn atomic column (**Fig. 1h**), showing that the dichroic signal peaks off-nucleus forming multi-pixel lobes, consistent with the anisotropic, delocalized nature of *d*-orbital transitions. To increase the dichroic contrast, we implement a two-window integration method (**Fig. 1i**), where $I_h$ and $I_v$ signals are obtained by sampling pairs of off-nucleus windows aligned with *d*-orbital antinodes ($I_h = L + R$; $I_v = T + B$). This approach selectively captures strong dipole-allowed transitions while excluding nodal regions, yielding enhanced ELD contrast compared with the line integration method (**Extended Data Fig. 2**).

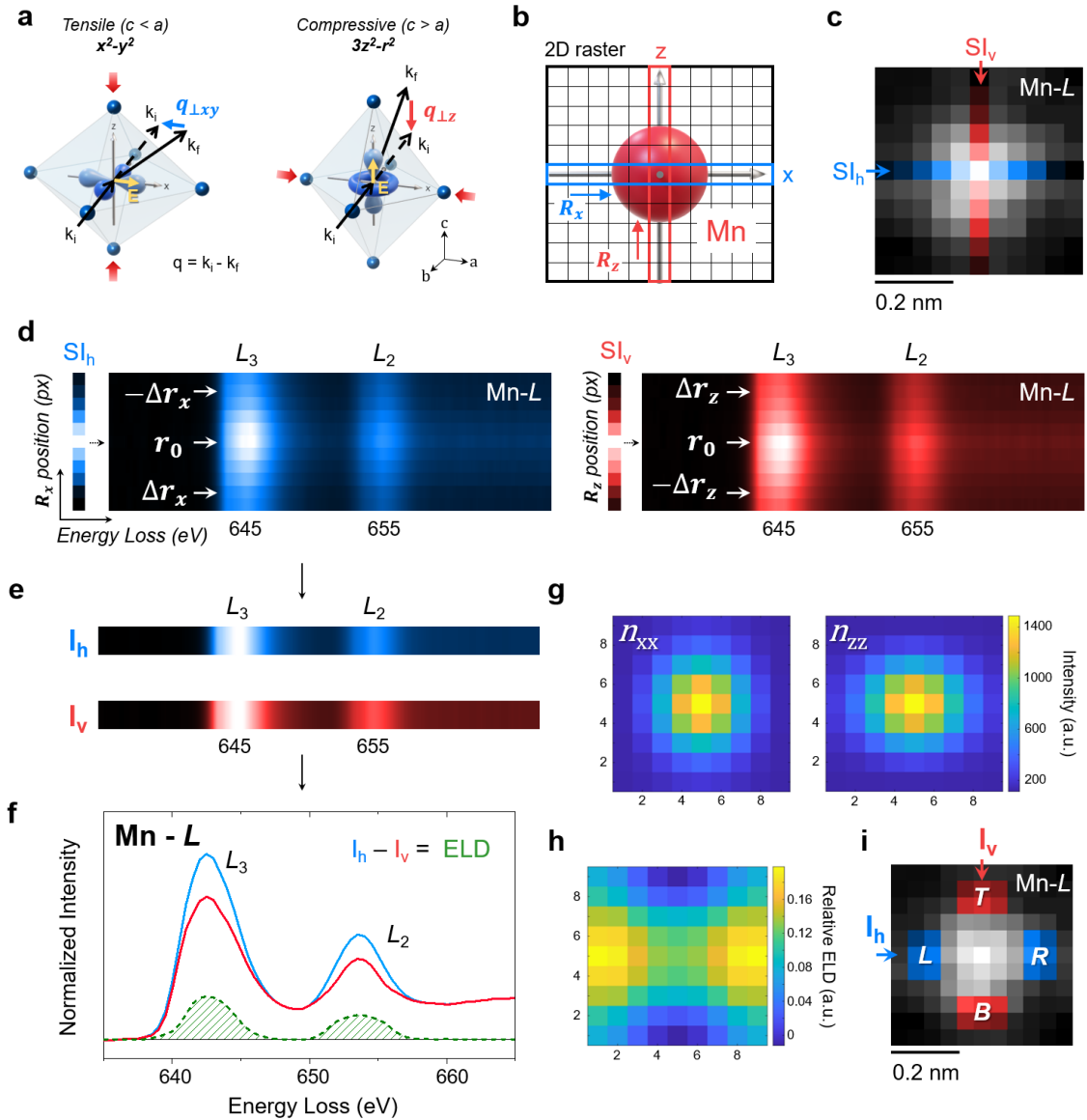

**Fig. 1 | Electron linear dichroism in the STEM. a,** Schematics of the Mn $3d$ $e_g$ orbitals ($x^2$-$y^2$ and $3z^2$-$r^2$) within MnO$_6$ octahedra, illustrating the scattering geometry, the corresponding momentum transfer vectors **q** and the strain-dependence orbital polarization, where the red arrows indicate the octahedra deformation direction. The diagrams show two perpendicular $\bm{q}_\perp$ main components: the in-plane $\bm{q}_{\perp x,y}$ and out-of-plane $\bm{q}_{\perp z}$ geometrically align with the $x^2$-$y^2$ and $3z^2$-$r^2$ orbitals, left and right panels respectively, which enables probing anisotropy in empty Mn $3d$ states. **b,** 2D STEM raster scan around the Mn atomic columns with the incident beam parallel to the [010] zone axis, and geometrically aligning with the orbitals in-plane ($x, y$) and out-of-plane ($z$) crystallographic axes. Each box (pixel) corresponds to a beam scan positions with in-plane, $\bm{R}_x$, and out-of-plane, $\bm{R}_z$, coordinates. For the ELD analysis, regions dominated by $\bm{q}_{\perp x}$ ($\bm{R}_x$) and $\bm{q}_{\perp z}$ ($\bm{R}_z$) are integrated along the $x$ and $z$ directions, crossing the center of the Mn atomic column, as indicated by the blue and red color boxes. **c,** Mn-$L$ energy-filtered spectrum image of a Mn column, showing in red and blue the integration area for the horizontal (SI$_h$) and vertical (SI$_v$) scan directions (line integration method). **d,** EELS spectra from the pixels along the two SI$_h$ and SI$_v$ scan directions. **e,** Total Mn-$L$ integrated intensity of $\bm{R}_x$ and $\bm{R}_z$ positions (pixels) along the two scan directions, I$_h$ and I$_v$ respectively. **f,** Illustrative ELD signal (I$_h$ – I$_v$) for a Mn$^{3+}$ ion in a tetragonal crystal field with (c > a). ELD gives information on the empty Mn-$3d$ states: high (low) absorption for I$_h$ indicates more in-plane (out of-plane) empty states in the $e_g$-band and thus a higher occupancy of out-of-plane (in-plane) orbitals. **g,** Dynamical diffraction calculations of Mn-$L$ energy-filtered images showing intensity maps of the spatially resolved, direction-dependent scattering coefficients $n_{xx}$ and $n_{zz}$, and **(h)** the their relative ELD signal for a Mn atomic column. These coefficients quantify the contribution of the dielectric response along the $x$- and $z$- crystal axes to the measured Mn-$L$ energy-filtered signal at each beam position, with their difference directly related to the ELD signal (see **Methods** for details). **i,** Two-window integration method, where I$_h$ = L + R and I$_v$ = T + B, targeting orbital antinodes to enhance dichroic contrast.

Building on the previously established strain–orbital coupling in LSMO, we performed monochromated ELNES experiments on epitaxial thin films grown on LaAlO$_3$ (LAO) and SrTiO$_3$ (STO) substrates, which impose compressive (c > a) and tensile (c < a) strain states into the LSMO, respectively. These measurements reproduce the known stabilization of $3d$–$e_g$ orbital occupancies —favoring $x^2$–$y^2$ under tensile strain and $3z^2$–$r^2$ under compressive strain— together with the expected XLD sign reversal[18,20,23]. Local strain mapping via geometrical phase analysis of high-resolution high-angle annular dark-field (HAADF) STEM images directly confirms the expected tetragonality changes in LSMO/LAO (**Fig. 2a**) and LSMO/STO (**Fig. 2b**).

**Fig. 2c** is a representative HAADF-STEM image acquired during spectrum imaging (SI), alongside its corresponding Mn-$L$ edge energy-filtered map. Using the procedure in **Fig. 1i**, I$_h$ and I$_v$ signals are obtained by integrating L + R (I$_h$) and T + B (I$_v$) off-nucleus windows, and the ELD signal is calculated as I$_h$ – I$_v$. For LSMO/LAO and LSMO/STO, the resulting ELD spectra reveal clear strain-dependent behavior (**Fig. 2e**). Under compressive strain (on LAO, c > a) the ELD signal is positive indicating I$_h$ > I$_v$, hence dominant $3z^2 – r^2$ occupation. Conversely, under tensile strain (on STO, c < a), the ELD signal is reversed, which demonstrates preferential $x^2 – y^2$ orbital filling.

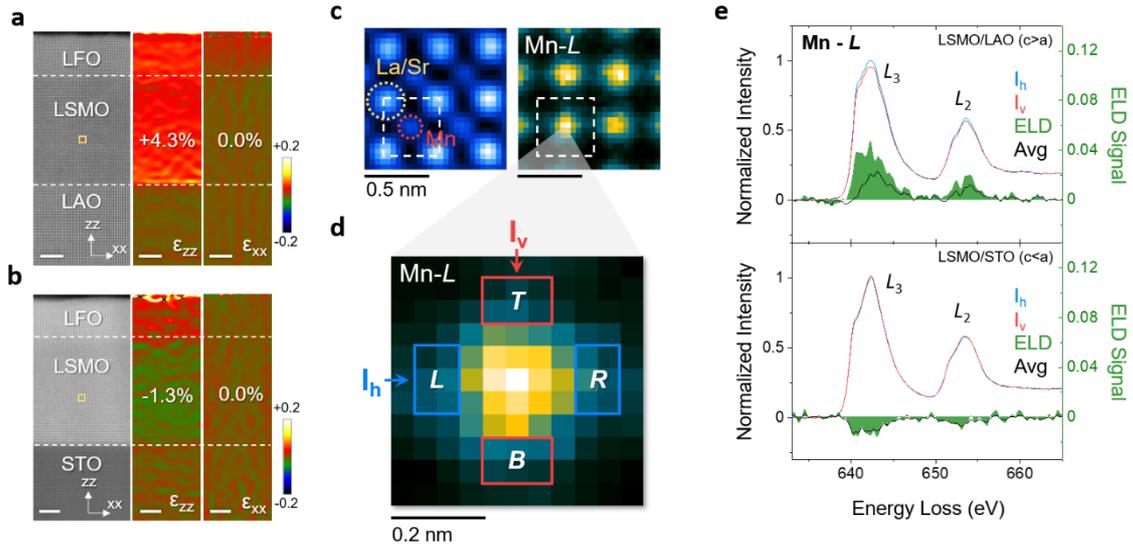

**Fig. 2 | Atomic-resolution mapping of orbital polarization of manganite thin films. a, b,** Left panels show atomic resolution HAADF-STEM images of epitaxial LFO/LSMO grown on LAO and STO substrates, respectively. The geometrical phase analysis (GPA) maps of the out-of-plane (middle panels) and in-plane (right panels), computed from the HAADF images in **a**, confirms that the LSMO is under compressive deformation (c>a) on LAO, while tensile (c<a) on STO. Labels on the GPA maps are the average strain value of LSMO relative to the reference lattice (substrate, zero strain). The square markers on the HAADF images represent the area where the EELS spectrum images were acquired for ELD. Scales bars are 5 nm. **c,** HAADF-STEM image of LSMO obtained during ELNES-SI acquisition (left) and its corresponding background-subtracted, Mn-$L$ energy-filtered map (right). The orange and red dashed circles mark the position of La/Sr and Mn columns, respectively. **d,** Mn-$L$ energy-filtered image of a single Mn column. The blue and red voxels indicate the 2x3-pixel integration windows used to extract the $I_h$ and $I_v$ signals, where $I_h = L + R$ and $I_v = T + B$, respectively. **e,** Normalized ELD calculated as $I_h - I_v$ from LSMO/LAO (top) and LSMO/STO (bottom) samples. The black curves indicate the average ELD signal across nine individual Mn atomic columns.

These measurements provide direct, atomic-scale verification of strain-controlled orbital polarization in manganites, in excellent agreement with prior XLD studies[19,23], confirming the validity of our experimental methodology and theoretical predictions of strain-orbital coupling. Our results also efficiently reproduce the orbital dependence on strain manifested by the weaker dichroic contrast for tensile LSMO compared to compressive, in good agreement with reports of a nearly linear dependence on the tetragonality[23]. The consistency of this strain-dependence is further reinforced by averaging the ELD signal across nine individual Mn atomic columns, as shown by the black curves in **Fig.2e**.

As a control experiment, we applied the same analysis to unstrained STO, examining the ELD of Ti-$L$ edge. In the case of cubic $TiO_6$ octahedra with degenerate $3d$ states in $O_h$ crystal field, we observed no detectable ELD signal (**Extended Data Fig. 3**), as expected for a set of $I_h$ and $I_v$ with identical absorption[24,25]. This null result confirms that our technique specifically detects symmetry-breaking orbital polarization.

While both XLD and ELD capture strain-induced spectral signatures linked to orbital occupancy, their spatial resolution capabilities differ. While XLD measurements typically average over micron-scale volumes, our STEM-ELD approach probes Ångström-scale volumes, enabling the detection of localized orbital distortions at single point defects and interfaces that are inaccessible to XLD.

The techniques also differ in their probing depths. XLD is inherently surface-sensitive, typically sampling only the top 5-10 nm of a sample, making measurements susceptible to artefacts of surface reconstruction or contamination. Our EELS measurements indicate a $Mn^{3+}$ fine structure (**Fig. 2e**) in contrast to XLD, where $Mn^{2+}$ fine structure is attributed to surface reduction and beam-stimulated oxygen desorption[20,26,27]. This is an advantage of STEM-ELD over XLD, as STEM-ELD can efficiently distinguish between local and bulk properties.

In conclusion, we have established a versatile STEM-based ELD methodology that enables atomic-resolution mapping of orbital polarization. The approach is compatible with any aberration-corrected microscope, requiring only zone-axis alignment and on-axis EELS geometry, and it should be universal for materials exhibiting in-plane symmetry breaking. The ELD approach employed here is largely insensitive to the use of a large collection aperture. This key advantage enables the acquisition of high-SNR ELNES spectra, even when using monochromated electron beams that achieve energy resolutions comparable to —or superior to— those of synchrotron X-rays spectroscopy. We introduced and validated two complementary ELD signal-extraction strategies, allowing users to select the optimal method based on experimental priorities such as pixel rate, dwell time optimization, SNR requirements, and sample stability. Although the ultimate spatial resolution is constrained by spectral SNR, the technique achieves single-atomic-column sensitivity and holds potential for single-atom orbital mapping detection. This capability unlocks unprecedent opportunities to probe orbital polarization and symmetry-breaking phenomena at point defects, interfaces, and in engineered quantum materials where conventional spectroscopic techniques fail to provide sufficient spatial resolution.

**Acknowledgments**

We acknowledge support by the CAS Project for Young Scientists in Basic Research (YSBR-003), the National Natural Science Foundation of China (grant No. 52373231), and the Beijing Outstanding Young Scientist Program (No. BJJWZYJH01201914430039). This research benefited from resources and support from the Electron Microscopy Center at the University of Chinese Academy of Sciences. The authors also acknowledge financial support from projects No.PID2020-118479RB-I00, No.PID2023-152225NB-I00 and No.PID2023-14947NB-I00, from Severo Ochoa MATRANS42 (No.CEX2023-001263-S) of the Spanish Ministry of Science, Innovation and Universities/AEI (Grant No. MICIU/AEI/10.13039/501100011033 and FEDER, EU), and from Generalitat de Catalunya (2021 SGR 00445). We also acknowledge the Swedish Research Council (grant no. 2021-03848), Olle Engkvist Foundation (grant no. 214-0331), STINT (grant no. CH2019-8211), Knut and Alice Wallenberg Foundation (grant no. 2022.0079) and eSSENCE for financial support. The simulations were enabled by resources provided by the National Academic Infrastructure for Supercomputing in Sweden (NAISS), partially financed by the Swedish Research Council through grant agreement no. 2022-06725. This work was partially supported by the U.S. Department of Energy (DOE), Office of Science (SC), Basic Energy Sciences, Material Sciences and Engineering Division, Electron and Scanning Probe Microcopies Program, FWP 83244 (JCI).

**METHODS**

**Sample growth**

Thin films of conducting $La_{0.7}Sr_{0.3}MnO_3$ (27 nm) were deposited by pulsed laser deposition (PLD) on STO and LAO single-crystalline substrates and capped with thin layers of $LaFeO_3$ (10 nm), following the growth conditions reported elsewhere[28].

**Scanning transmission electron microscopy**

Samples for cross-sectional investigations were prepared by standard lift-out procedures using a Thermo Fisher Scientific HELIOS dual-beam focused ion beam system. Protective Pt layers were applied over the region of interest before cutting and milling. To minimize the sidewall damage and ensure a sufficiently thin specimen for electron transparency, final milling was carried out at a voltage of 2 kV. Atomic-resolution high angle annular dark-field (HAADF) images of the LFO/LSMO/substrate interfaces show in in **Fig. 2a-b** were carried out in a JEOL GrandARM-2 operated at 200 kV. The collection semi-angles were set to 68-280 mrad. To minimize distortions from the scanning system for strain mapping quantification, HAADF images were acquired sequentially with short dwell time at 0º and 90º respect to the interface plane. The 0º and 90º image stacks were integrated after drift-correction, and only the strain component along the fast-scan direction was used for strain quantification, respectively. Strain maps were computed using the geometrical phase analysis (GPA) by masking the (100) and (001) reflections, and defining the substrate as the reference, unstrained lattice.

**Electron linear dichroism**

Monochromatic energy loss near edge structure (ELNES) spectroscopy was performed in a NION U-HERMES 100 microscope operated at 60 kV, equipped with a NION monochromator and a Dectris ELA hybrid-pixel detector, specifically optimized for electron energy-loss spectroscopy (EELS). The probe convergence semi-angle was set to 32 mrad. ELNES spectra were acquired in standard on-axis geometry with no spectrometer aperture inserted, which gives a collection semi-angle of 75 mrad, set by the HAADF inner collection semi-angle (75-210). After monochromation, the beam current was 15-20 pA, which ensured a stable energy resolution ~50 meV with the beam intersecting the sample. Multi-frame ELNES spectrum images were recorded using a dispersion of 0.05 eV/channel, using a two-dimensional scan window of 2.57 x 2.57 nm (~ 4 x 4 LSMO unit cells). The scan width was set to 50 x 50 pixels (~ 51 pm/pixel), with a dwell time of 50 ms/frame with a total of 10 frames. We found this set up to be the best compromise between pixel-rate per column, SNR and minimum sample drift. The averaged thickness of both samples was between 25 and 30 nm, as determined by the measurement of the zero-loss peak.

After acquisition, the stack of spectrum images was aligned and integrated using a cross-correlation method. Spectrum images were denoised using principal component analysis, and reconstructed using the first three principal components, and then a gaussian convolution was applied using an in-house python code based on Numpy and Scipy modules to further smooth the spectra ensuring minimum broadening of the fine structure spectral shape. Spectrum images were background subtracted by fitting a power law function in the pre-edge energy range. For the "line-integration method", two 7 x 1-pixel windows were used to integrate the signal along the horizontal and vertical scan directions across the center of the Mn columns to obtain the horizontal ($I_h$) and the vertical ($I_v$) signals, respectively, as indicated in **Fig. 1c**. In the case of the "two-window integration method", two pairs of 2 x 3 off-nucleus windows were placed aligned with $d$-orbital antinodes to obtain $I_h$ and $I_v$, as illustrated in **Fig. 1i**. After summation, $I_h$ and $I_v$ spectra were normalized to the maximum intensity at the energy loss of 642 eV, and then the $I_v$ signal was adjusted to match the one of the $I_h$ in the Mn-$L_2$ post-edge energy range, ensuring zero ELD signal after subtraction ($I_h - I_v$).

**Computational details**

For non-magnetic materials with a tetragonal structure, the dielectric tensor is diagonal with two equal elements along the $a = b$ crystal axes and a distinct one along the $c$-axis[29]. The difference between these two diagonal elements is proportional to the linear dichroism (LD):

$$\text{LD}(E) \propto \epsilon_c(E) - \frac{1}{2}[\epsilon_a(E) + \epsilon_b(E)]. \quad (1)$$

Since $\epsilon_a(E) = \epsilon_b(E)$, this simplifies to:

$$\text{LD}(E) \propto \epsilon_c(E) - \epsilon_a(E). \quad (2)$$

Defining the isotropic average diagonal element:

$$\epsilon(E) = \frac{1}{3}[\epsilon_a(E) + \epsilon_b(E) + \epsilon_c(E)], \quad (3)$$

and noting that $\epsilon_a(E) = \epsilon_b(E)$, the individual diagonal elements can be written as:

$$\epsilon_a(E) = \epsilon_b(E) = \epsilon(E) - \frac{1}{3}\text{LD}(E),$$

$$\epsilon_c(E) = \epsilon(E) + \frac{2}{3}\text{LD}(E). \quad (4)$$

For the EELS calculations, we take the electron beam parallel to the $b$-axis, with scan directions $(x, z)$ aligned to the crystal axes $(a, c)$. The spectrum at a given beam position $(x, z)$ is:

$$\sigma(x, z, E) = \epsilon_a(E)n_{xx}(x, z) + \epsilon_b(E)n_{yy}(x, z) + \epsilon_c(E)n_{zz}(x, z), \quad (5)$$

where $n_{xx}$, $n_{yy}$ and $n_{zz}$ are the geometrical projected scattering coefficients determined by convergence and collection angles, crystal structure, and sample thickness. Using the expressions above we can rewrite the spectrum as:

$$\sigma(x,z,E) = \epsilon(E)[n_{xx}(x,z) + n_{yy}(x,z) + n_{zz}(x,z)] + \frac{1}{3}\text{LD}(E)[2n_{zz}(x,z) - n_{xx}(x,z) - n_{yy}(x,z)], \quad (6)$$

Note that for tetragonal symmetry with the incoming electron beam along $b$-axis, we expect:

$$n_{xx}(x, z) = n_{xx}(\pm x, \pm z), \quad (7)$$

and similar for other components $n_{zz}(x, z)$. However, $n_{xx}(x, z)$ should differ from $n_{xx}(z, x)$ due to $q_x$-selective terms in mixed dynamical form factors (MDFFs). This yields nonzero differences in the maps of $n_{xx}(x, z) - n_{xx}(z, x)$ and similarly for $n_{zz}(x, z) - n_{zz}(z, x)$ shown in in **Fig. 1g**.

In actual experiments, we sum the spectra from sets of $x,z$ electron beam positions and then perform a post-edge normalization. Since $LD(E)$ has zero intensity in the post-edge region, the normalization factors are proportional to the prefactor of isotropic spectrum $\epsilon(E)$ in Eq. (6). To get a normalized spectrum, we thus need to divide it by a constant times $n_{xx}(x,z) + n_{yy}(x,z) + n_{zz}(x,z)$ to obtain

$$\sigma_{\text{norm}}(x,z,E) = \epsilon(E) + \frac{1}{3} LD(E) \frac{2n_{zz}(x,z) - n_{xx}(x,z) - n_{yy}(x,z)}{n_{xx}(x,z) + n_{yy}(x,z) + n_{zz}(x,z)}, (8)$$

The factor multiplying the *LD*(*E*) is the map shown in **Fig. 1h**. When we aim to measure linear dichroism *LD*(*E*), we need to pick two sets of beam positions where this map shows large differences while maximizing signal to noise ratio, as is illustrated in **Fig. 1i**. Taking difference of two spectra will cancel out the isotropic spectrum $\epsilon(E)$ and leave a result proportional to the *LD*(*E*).

Finally, we note that magnetic materials will have a non-diagonal dielectric tensor due to presence of electron magnetic circular dichroism (EMCD). However, as long as the magnetic moment points along one of the lattice vectors, the method still works thanks to large effective detector collection angle and because we are summing spectra over pixels connected by mirror symmetries, which flip the sign of the EMCD signal. These symmetries will then cause mutual cancellation of the EMCD intensity in the sums over beam positions, as indicated in **Fig. 1i**.

**Data availability**

The data that support the findings of this study are available from the corresponding authors on request.

**Code availability**

All of the codes used in this work are available from the corresponding authors on request.

**Contributions**

J.G., J.C.I. and J.R. conceived the project; J.G, J.C.I. and W.Z. supervised the research; R.G. and W.Z. designed the research and microscopy experiments with input from J.G. and J.C.I.; R.G. prepared the TEM specimens, performed the STEM experiments and data processing under the supervision of W.Z.; A. L. wrote support codes for spectral data processing; J.R. designed and carried out the dynamical diffraction calculations. The manuscript was prepared by R.G. with contributions from all other co-authors.

**Ethics declarations**

Competing interests

The authors declare no competing interests.

**EXTENDED DATA**

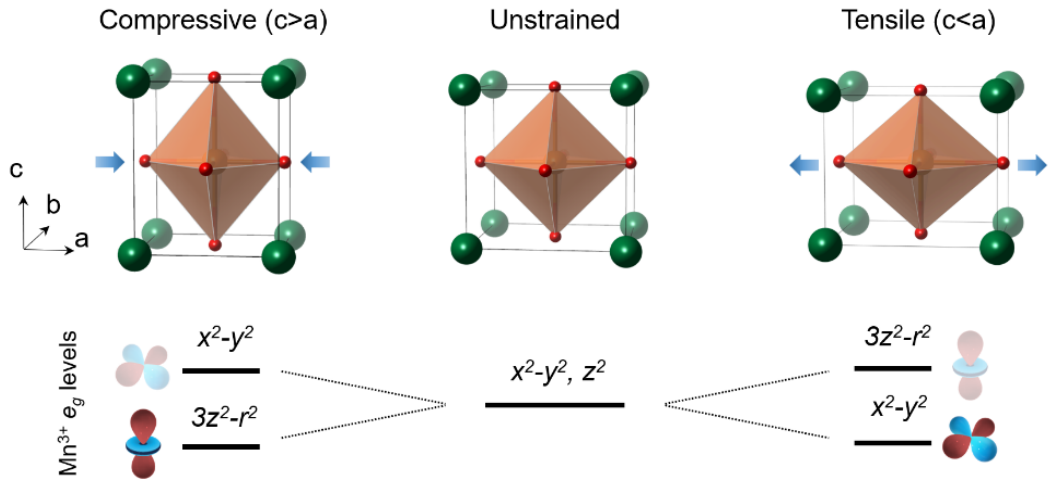

**Extended Data Fig. 1 | Strain-induced orbital polarization in LSMO.** Top panels: $MnO_6$ octahedral distortions as a function of strain for (001)-oriented LSMO, where green, orange and red spheres represent La/Sr, Mn and O, respectively. Bottom panels: the effects of symmetry breaking on the Mn $e_g$ levels of $Mn^{3+}$ ions. Strain induces a tetragonal crystal field in which compressive strain (c > a) favors $3z^2-r^2$ occupancy whereas tensile strain (c < a) favors $x^2-y^2$ occupancy.

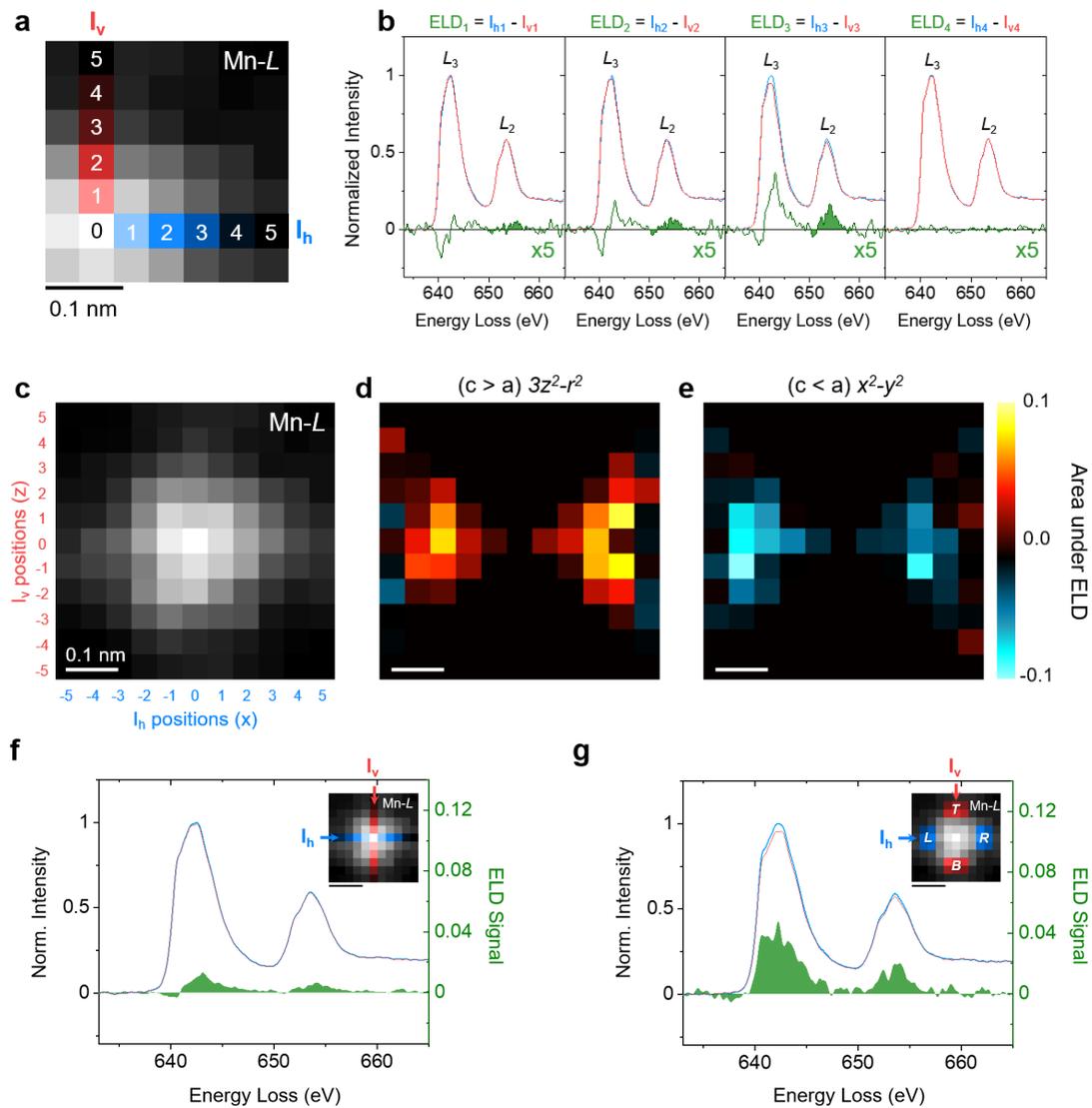

**Extended Data Fig. 2 | Atomic-resolution electron linear dichroism. a,** Quadrant of a Mn-$L_{2,3}$ energy-filtered spectrum image, indicating horizontal ($I_h$, blue) and vertical ($I_v$, red) integration regions for the line-integration method. Pixel numbering begins at the Mn column center (0). **b,** Position-dependent ELD signal peaking ~3 pixels from the column center (green shading: Mn-$L_2$ integration window). **c,** Coordinate map of the Mn-$L$ spectrum image. **d, e,** Orbital polarization maps for compressively strained (LAO, **d**) and tensile strained (STO, **e**) LSMO, showing sign reversal of the area under ELD[19,23], with maxima displaced from atomic nuclei. Color denotes the integrated ELD value at each beam position, calculated as the difference between horizontal ($I_{h(x)}$) and vertical ($I_{v(z)}$) coordinate positions at each pixel. Non-integrated pixels are set to zero (black). **f, g,** Comparison of the line-integration (**f**) and two-window integration (**g**) methods for the same dataset, demonstrating enhanced contrast with the latter. The scale bars in the insets are 0.1 nm.

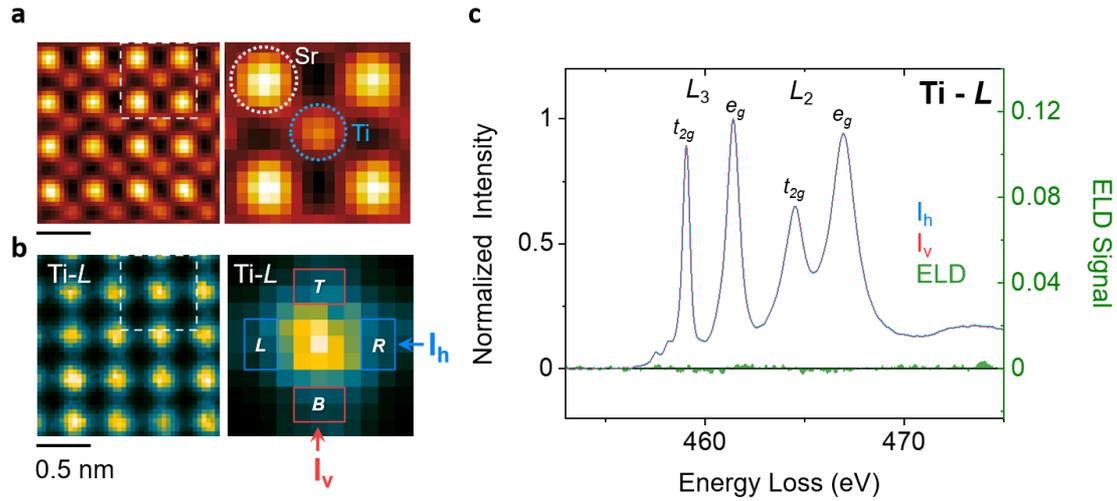

**Extended Data Fig. 3 | Electron linear dichroism of unstrained SrTiO₃. a,** left: HAADF-STEM image of unstrained SrTiO₃, acquired during ELNES-SI collection. Right: magnified view of a single SrTiO₃ unit cell, with white and blue dashed circles marking Sr and Ti atoms, respectively. **b,** Left: corresponding Ti-$L$ energy filtered map to the image in **a**. Right: magnified unit cell with blue and red markers indicating the 2 x 3-pixel integration windows used to extract the $I_h$ and $I_v$ signals. **c,** Normalized ELNES spectra for in-plane ($I_h$= L+R, blue) and out-of-plane ($I_v$=T+B, red) orientations in bulk-SrTiO₃. The ELD signal $I_h – I_v$ is negligible, as expected for Ti-3$d$ orbitals under a cubic crystal field.